\documentclass[aip,chaos,reprint]{revtex4-1}
\usepackage[usenames,dvipsnames]{color}
\usepackage{graphicx}
\usepackage{amsmath}

\begin{document}
\title[]{Giant oscillations of diffusion in ac-driven periodic systems}
\author{I. G. Marchenko}
\affiliation{NSC \lq\lq Kharkov Institute of Physics and Technology\rq\rq, Kharkov 61108, Ukraine}
\affiliation{Kharkov National University, Kharkov 61077, Ukraine}
\affiliation{Institute of Physics, University of Silesia, 41-500 Chorz{\'o}w, Poland}

\author{A. Zhiglo}
\affiliation{NSC \lq\lq Kharkov Institute of Physics and Technology\rq\rq, Kharkov 61108, Ukraine}
\affiliation{Max Planck Institute for Astronomy, 69117 Heidelberg, Germany}

\author{V. Aksenova}
\affiliation{NSC \lq\lq Kharkov Institute of Physics and Technology\rq\rq, Kharkov 61108, Ukraine}
\affiliation{Kharkov National University, Kharkov 61077, Ukraine}

\author{V. Tkachenko}
\affiliation{NSC \lq\lq Kharkov Institute of Physics and Technology\rq\rq, Kharkov 61108, Ukraine}
\affiliation{Kharkov National University, Kharkov 61077, Ukraine}

\author{I. I. Marchenko}
\affiliation{NTU \lq\lq Kharkov Polytechnic Institute\rq\rq, Kharkov 61145, Ukraine}

\author{\\J. {\L}uczka}
\author{J. Spiechowicz}
\email{jakub.spiechowicz@us.edu.pl}
\affiliation{Institute of Physics, University of Silesia, 41-500 Chorz{\'o}w, Poland}

\begin{abstract}
We revisit the problem of diffusion in a driven system consisting of an inertial Brownian particle moving in a symmetric periodic potential and subjected to a symmetric time-periodic force. We reveal  parameter domains in which diffusion is normal in the long time limit and exhibits intriguing giant damped quasiperiodic oscillations as a function of the external driving amplitude. As the mechanism behind this effect we identify the corresponding oscillations of difference in the number of locked and running trajectories which carries the leading contribution to the diffusion coefficient. Our findings can be verified experimentally in a multitude of physical systems including colloidal particles, Josephson junction or cold atoms dwelling in optical lattices, to name only a few.
%An inertial Brownian particle moving in a symmetric periodic potential and driven by a symmetric time-periodic force of amplitude $A$ can exhibit a huge damped oscillations of the diffusion coefficient as a function of $A$. We explore the parameter space of the system and find the regimes of high-frequency driving in which giant quasi-periodic diffusion can occur. We identify the primary mechanism responsible for the phenomenon: it is deterministic dynamics of running and locked states, and difference in populations of both states perturbed by thermal fluctuations.  
\end{abstract}
\maketitle
\begin{quotation}
Processes occurring at the micro and nano-world are strongly influenced by thermal fluctuations whose intensity is determined by temperature of the system. A typical example is Brownian motion of a classical particle with erratic and unpredictable trajectories that are commonly characterized by a diffusion coefficient which in equilibrium is proportional to temperature. Because the ac-driven systems are out of equilibrium there can exist certain domains of the system parameters in which unexpected behaviour is encountered.  In this work, we study a paradigmatic model of diffusion phenomena in nonequilibrium statistical physics that for decades has been applied to classical systems. We identify  parameter regimes in which diffusion is normal in the long time limit, however, its dependence on the system parameter exhibits intriguing giant damped quasi-periodic oscillations. Our findings can be corroborated experimentally in various systems including, among others, colloidal systems, Josephson junctions and cold atoms dwelling in optical lattices.
\end{quotation}

\section{Introduction}
Processes and phenomena in systems far from equilibrium can exhibit unexpected and unusual  properties. They are ubiquitous in Nature ranging from the microscale to the macroscale. 
We quote only a few simplest and popular examples intensively studied in physics: stochastic \cite{benzi1981,gammaitoni1998} or coherence \cite{pikovsky1997, lindner2004} resonance, ratchet effect \cite{hanggi2009, actrat2017}, absolute negative mobility \cite{machura2007,nagel2008,slapik2019,spiechowicz2019njp}, anomalous transport and anomalous diffusion \cite{klages2008, metzler2014, marchenko2014,
spiechowicz2016njp, marchenko2017,spiechowicz2017chaos, marchenko2018,marchenko2019}.
Diffusion plays a key role in numerous processes in physics, chemistry, biology and engineering. Moreover, this concept is exploited in sociology, politics and culture as a measure of spread of ideas, values, concepts, knowledge, practices, behaviors, materials, symbols and so on
\cite{socjology}. 
 For normal diffusion described by the Einstein relation for a Brownian particle, the diffusion coefficient $D$ depends only on two parameters: it monotonically  increases with increasing temperature $T$ and decreases when the friction coefficient grows. 
 However, under nonequilibrium conditions, the dependence of $D$ on the system parameters may be surprising and non-monotonic. There are many scenarios to move systems out of equilibrium. One of the simplest way is to apply a constant force. The next is to use a time periodic perturbation. Recently, the great interest has been attracted to active fluctuations \cite{marchetti2013, spiechowicz2014pre, bechinger2016} which appear e.g. in living cells where energy is provided in the form of  biochemical reactions that drives active cellular processes \cite{gnesotto2018}. In such systems diffusion can be normal, however even then it is usually non-Gaussian \cite{wang2012, metzler2017, bialas2020} or it simply is anomalous \cite{metzler2014}, i.e. exhibits subdiffusive or superdiffusive behaviour. In the paper, we consider the case when the applied time periodic force continually supplies energy  to the system and drives it   far from thermodynamic equilibrium.  We study diffusion and its dependence on the amplitude of this external driving. We find parameter regimes where the diffusion coefficient exhibits intriguing giant damped quasi-periodic oscillations.

The paper is organized as follows. In Sec. II, we formulated the model in terms of the Langevin equation. In Sec. III, we define quantifiers used to characterize the diffusion process. Sec. IV contains a description of simulations. In Sec. V we present the main result of the paper, i.e. the dependence of the diffusion coefficient on the amplitude of the ac-driving and explain its origin. The last section provides  summary.

\section{Description of the model}

We study a relatively simple model of a system in a nonequilibrium state. It is a classical Brownian particle of mass $M$ subjected to a one-dimensional, spatially periodic potential $U(x)$ and driven by an unbiased and symmetric time-periodic force $F(t)$. Its dynamics can be described by the Langevin equation in the form \cite{spiechowicz2016njp}
\begin{equation}
	\label{model}
	M\ddot{x} + \Gamma\dot{x} = -U'(x) + F(t) + \sqrt{2\Gamma k_B T}\,\xi(t),
\end{equation}
where the dot and the prime denotes  differentiation with respect to time $t$ and the particle coordinate $x$, respectively. The parameter $\Gamma$ is the friction coefficient. The potential $U(x)$ is assumed to be symmetric of spatial period $L$ and the barrier height  $2 \Delta U$ reading
\begin{equation}
	\label{potential}
	U(x) = U(x+L) = -\Delta U\cos{\left( \frac{2\pi}{L}x \right)}.
\end{equation}
The external ac-driving force of amplitude $A$ and angular frequency $\Omega$ has the simplest harmonic form  
\begin{equation}
	F(t) = A \sin{(\Omega t)}. 
\end{equation}
Thermal equilibrium fluctuations due to interaction of the particle with its environment of temperature $T$ are modelled as $\delta$-correlated Gaussian white noise of zero-mean value, 
\begin{equation}
	\langle \xi(t) \rangle = 0, \quad \langle \xi(t)\xi(s) \rangle = \delta(t - s), 
\end{equation}
where the bracket $\langle \cdot \rangle$ denotes an average over white noise realizations (ensemble average). 
The noise intensity $2\Gamma k_B T$ in Eq. (\ref{model}) follows from the fluctuation-dissipation theorem \cite{kubo1966}, where $k_B$ is the Boltzmann constant. If $A=0$ the stationary state is a thermal equilibrium state. If $A \neq 0$, then the external force $F(t)$ drives the system away from the equilibrium state. 

There are several physical systems \cite{risken} that can be modelled by Eq. (\ref{model}). One can mention  the semiclassical dynamics of the phase difference across a Josephson junction and its variations including e.g. the SQUIDs \cite{kautz1996, spiechowicz2015chaos, blackburn2016} as well as the dynamics of cold atoms dwelling in optical lattices \cite{denisov2014,lutz2013,kindermann2017}. %or the transport of ions in superionic conductors \cite{fulde1975,dieterich1980}.

The model given by Eq. (1) has been studied for decades and used for analysis of 
various noisy phenomena in both deterministic regular and chaotic regimes. We refer the 
interested reader to the review paper \cite{kautz1996} and references therein. Yet, 
there still remain new phenomena to be uncovered for this system, which in turn carry 
the potential for new applications. As recent examples we can quote a non-monotonic 
temperature dependence of a diffusion coefficient in normal diffusion regimes 
\cite{spiechowicz2016njp, spiechowicz2017chaos, marchenko2018} and transient, yet 
extended time-dependent anomalous diffusion \cite{spiechowicz2016scirep, 
spiechowicz2017scirep, spiechowicz2019chaos}, to name but a few.

Now, we transform Eq. (1) to its dimensionless form. To this aim we use the following scales as  characteristic units of length and time
\begin{equation}
	\label{scaling}
	\hat{x} = 2\pi \frac{x}{L}, \quad \hat{t} = \frac{t}{\tau_0}, \quad \tau_0 = \frac{L}{2\pi}\sqrt{\frac{M}{\Delta U}}.
\end{equation}
Under such a procedure Eq. (\ref{model}) assumes the form
\begin{equation}
	\label{dimless-model}
	\ddot{\hat{x}} + \gamma\dot{\hat{x}} = -\sin{\hat{x}} + a \sin (\omega \hat{t}) +  \sqrt{2\gamma Q} \hat{\xi}(\hat{t}).
\end{equation}
In this scaling the dimensionless mass $m = 1$ and the remaining four dimensionless parameters  read
\begin{eqnarray}
	\gamma = \frac{\tau_0}{\tau_1}, \quad a = \frac{1}{2\pi}\frac{L}{\Delta U} A,  
\quad \omega = \tau_0 \Omega, \quad Q = \frac{k_B T}{\Delta U}, 
\end{eqnarray}
where the second characteristic time  is $\tau_1 = M/\Gamma$. It has the physical interpretation of the relaxation time for the velocity of the free Brownian particle. On the other hand, the characteristic time  $\tau_0$ is related to the period of small  oscillations inside the potential $U(x)$ wells. 

The rescaled potential of the period $L=2\pi$ is $\hat{U}(\hat{x}) = U((L/2\pi)\hat{x})/\Delta U = -\cos{\hat x}$ and the corresponding potential force is $-\hat{U}'(\hat{x})=-\sin \hat{x}$. The rescaled thermal noise is $\hat{\xi}(\hat{t}) = (L/2\pi \Delta U)\xi(t) = (L/2\pi \Delta U)\xi(\tau_0\hat{t})$ and has the same statistical properties as $\xi(t)$; i.e., $\langle \hat{\xi}(\hat{t}) \rangle = 0$ and $\langle \hat{\xi}(\hat{t})\hat{\xi}(\hat{s}) \rangle = \delta(\hat{t} - \hat{s})$. The dimensionless noise intensity $Q$ is the ratio of thermal energy and half of the activation energy the particle needs to overcome the nonrescaled potential barrier. From now on we shall stick to these dimensionless variables. In order to simplify the notation further we omit the hat-notation in Eq. (\ref{dimless-model}).
\begin{figure}[t]
\centering
\includegraphics[width=0.9\linewidth]{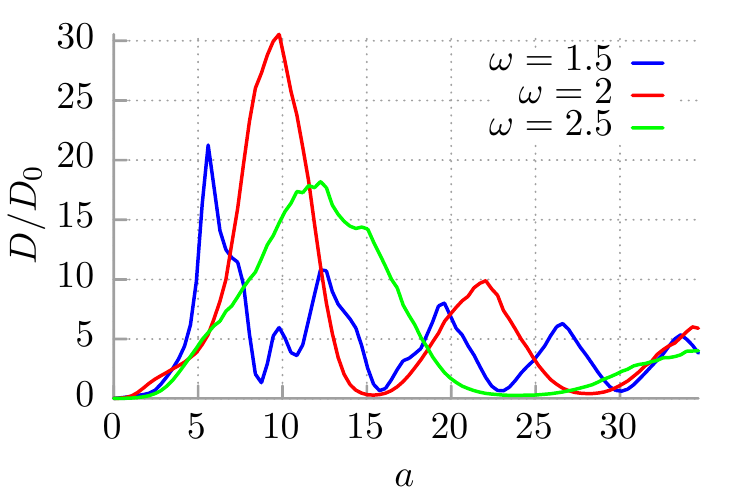}
\caption{The rescaled diffusion coefficient $D/D_0$, where \mbox{$D_0 = Q/\gamma$} is free thermal diffusion,  as a function of the  amplitude $a$ of the ac-driving for selected values of its frequency $\omega$. Other parameters are: $\gamma = 0.03$ and temperature $Q=0.5$} %Inset: $D_{max}(F_e)$  at $\omega=1.59$. The power law fit $D_{max} (F_e)={D}F_e^{-0.85}$ is shown as a dashed line.}
\label{fig1}
\end{figure}

\section{Quantifiers used for analysis of diffusion}

Diffusion is characterized by the mean square deviation (variance) of the particle position $x(t)$, namely,
\begin{equation}
	\label{msd}
	\langle \Delta x^2(t) \rangle = \langle \left[x(t) - \langle x(t) \rangle \right]^2 \rangle = \langle x^2(t) \rangle - \langle x(t) \rangle^2.
\end{equation}
Here and below $\langle \cdot \rangle$ stands for the average over thermal noise realizations as well as  over initial coordinates $x(0)$ and velocities $v(0)=\dot{x}(0)$ of the Brownian particle.
For normal diffusion regime $\langle \Delta x^2(t) \rangle$  is a linear function of time and the diffusion coefficient can be defined by the relation \cite{spiechowicz2016scirep}
\begin{equation}
	\label{diffusioncoefficient}
	D =  \lim_{t \to \infty} \frac{\langle \Delta x^2(t) \rangle}{2t}.
\end{equation}

As we will demonstrate the spread of position trajectories measured by the diffusion coefficient can be determined by behavior of the particle velocity. A relevant quantity characterizing transport of the Brownian particle is time averaged velocity
\begin{equation}
	\overline{v} = \lim_{t \to \infty} \frac{1}{t} \int_0^t ds \, \dot{x}(s). 
\end{equation}
Since the system given by Eq. (\ref{dimless-model}) is symmetric and unbiased a preferential direction of motion is forbidden in the long time limit. Consequently, the ensemble and time averaged velocity must vanish for both zero and non-zero temperature regimes \cite{denisov2014}
\begin{equation}
	\langle \overline{v} \rangle \equiv 0.
\end{equation}
 The latter is mandatory for the deterministic variant of dynamics which may be non-ergodic and therefore sensitive to the specific choice of starting position and velocity of the particle \cite{spiechowicz2016scirep}. However,   for any non-zero temperature $Q>0$  the system is ergodic and initial conditions do not affect properties of the system in the long time stationary regime.  Due to the above condition the spread of time averaged velocity can be quantified solely by its second moment, i.e.
\begin{equation}
	\sigma_{\overline{v}}^2 = \langle \overline{v}^2 \rangle - \langle \overline{v} \rangle^2 \equiv \langle \overline{v}^2 \rangle.
\end{equation}
\begin{figure}[t]
\centering
\includegraphics[width=0.9\linewidth]{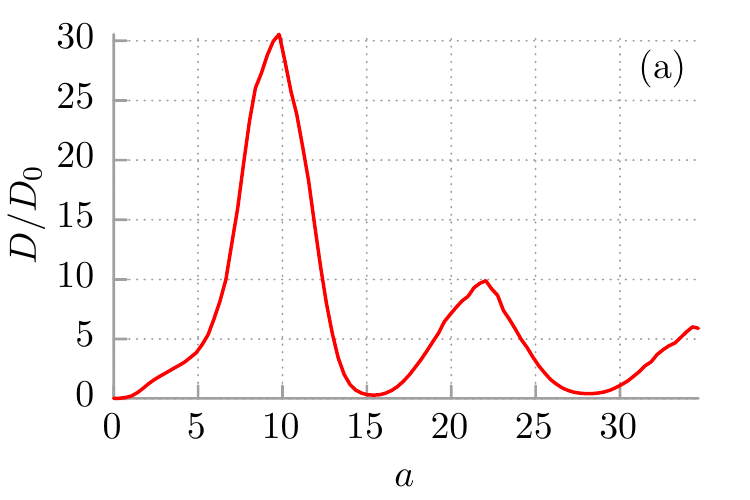}
\includegraphics[width=0.9\linewidth]{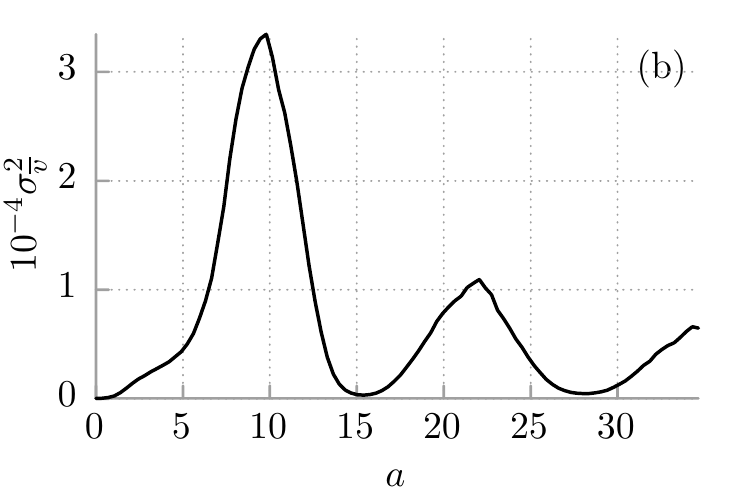}
\caption{The rescaled diffusion coefficient $D/D_0$ (panel (a)) and the variance of the time averaged velocity $\sigma_{\overline{v}}^2$ (panel (b)) for the fixed frequency $\omega =2$. Other parameters are the same as in Fig. \ref{fig1}.}
\label{fig2}
\end{figure}

\begin{figure*}[t]
\centering
\includegraphics[width=0.45\linewidth]{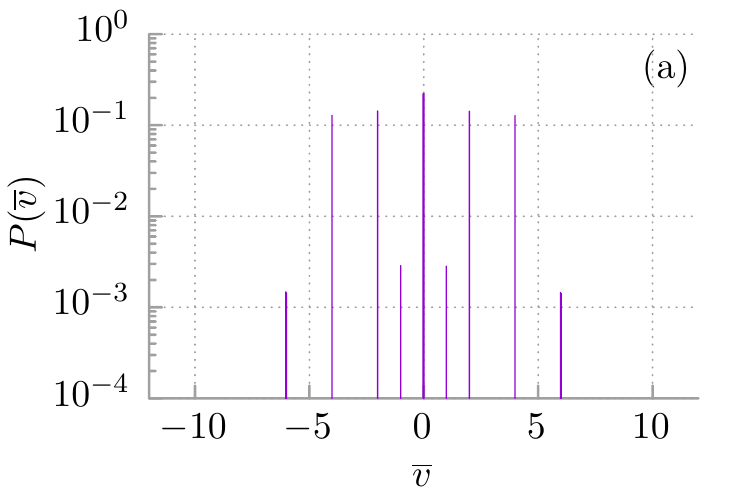}
\includegraphics[width=0.45\linewidth]{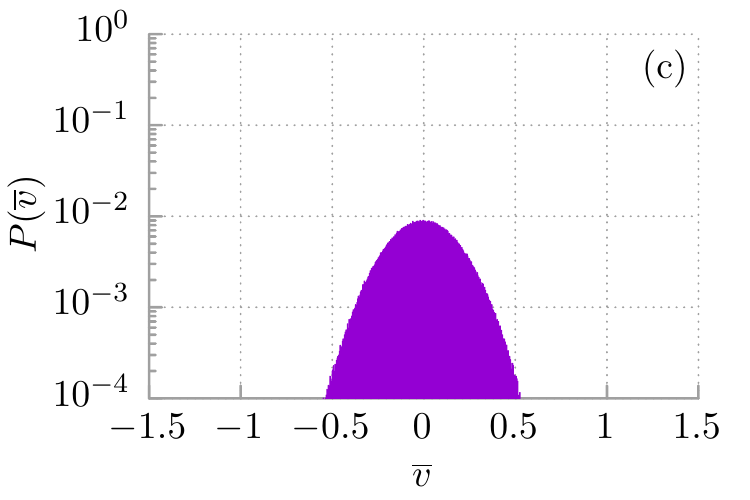}\\
\includegraphics[width=0.45\linewidth]{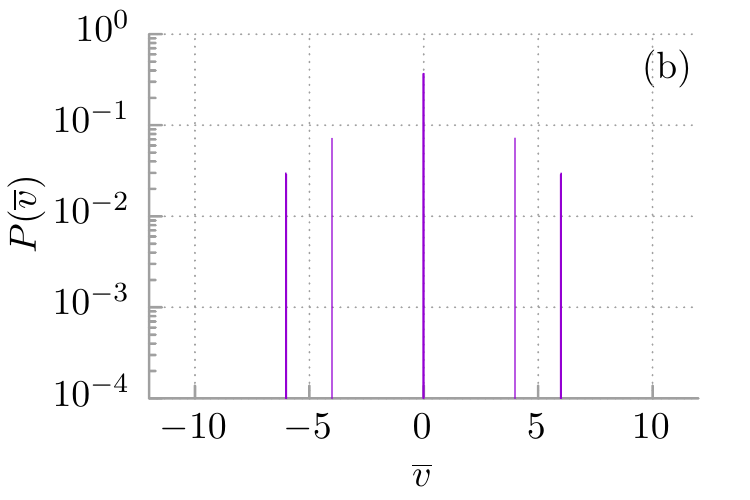}
\includegraphics[width=0.45\linewidth]{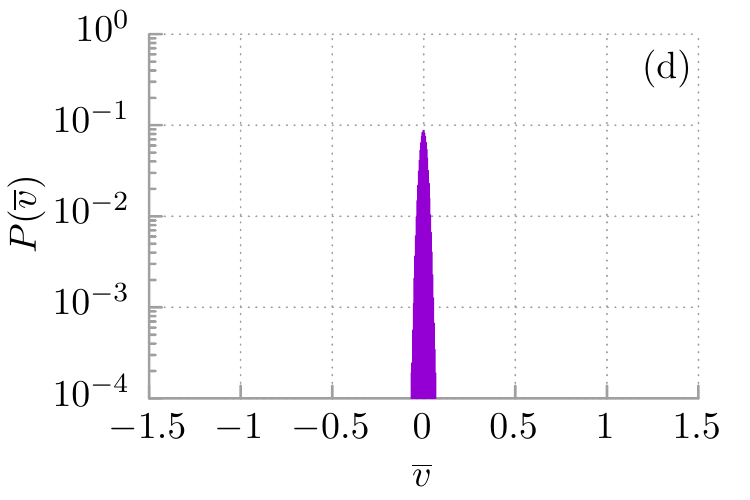}
\caption{The probability distributions $P(\overline{v})$ for the time averaged velocity $\overline{v}$ of the Brownian particle in the deterministic limit of vanishing temperature $Q = 0$ and noisy dynamics $Q = 0.5$ are depicted in left and right panels, respectively. Upper plots ((a) and (c)) represent the first maximum ($a = 9.8$) in diffusion oscillations whereas bottom ones ((b) and (d)) correspond to the first minimum ($a = 15.3$). Other parameters are the same as in Fig. \ref{fig2}.}
\label{fig3}
\end{figure*}

\section{Description of the simulations}

The complexity of stochastic dynamics determined by Eq. (\ref{dimless-model}) with three-dimensional phase space  $\{x, y=\dot x, z=\omega t\}$ is rooted in the four-dimensional parameter space $\{\gamma, a, \omega, Q\}$.  The  Fokker-Planck equation corresponding to Eq. (\ref{dimless-model}) cannot be solved analytically and for this reason we had to resort to comprehensive numerical simulations. All calculations have been done using a Compute Unified Device Architecture (CUDA) environment implemented on a modern desktop Graphics Processing Unit (GPU). This proceeding allowed for a speedup of factor of the order $10^3$ times as compared to present day Central Processing Unit (CPU) method \cite{spiechowicz2015cpc}. The Langevin equation (\ref{dimless-model}) were integrated using a second order predictor-corrector scheme \cite{platen} with the time step $h = 10^{-2}$. The quantities characterizing diffusive behaviour of the system were averaged over the ensemble of $N = 2^{14} = 16384$ trajectories, each starting with different initial conditions $x(0)$ and $v(0)$ distributed uniformly over the intervals $[0, 2\pi]$ and $[-2,2]$, respectively. Under the assumption of asymptotic normality of the estimator of the mean of the quantities characterizing diffusive behavior of our system, we obtain that the statistical error of the Monte Carlo simulation is of the order of $\approx 10^{-2}$, which seems adequate for our purposes. The time span of simulations read $10^6$ periods $2\pi/\omega$ of the external driving $a\cos{(\omega t)}$ and was extensive enough to reach the long time limit indicated by the stationarity of diffusion coefficient.

\section{Oscillations of the diffusion coefficient}

The spatially periodic potential and the ac-driving are reflection symmetric with respect to coordinate and time, respectively. Therefore in the long time limit the averaged particle velocity is zero $\langle \overline{v} \rangle = 0$ and no directed transport can be generated in the studied system. Moreover, for non-zero temperature $Q>0$ the diffusion is normal in the long time limit and therefore its coefficient $D$ is well-defined (i.e. it has a finite value larger than zero). 
In some regimes of the parameter space $D$ can render a non-trivial behavior. E.g. the Einstein relation fails and non-monotonic temperature dependence of the diffusion coefficient can be detected. Because the ac-driving takes the system out of thermal equilibrium the influence of its parameters $a$ and $\omega$ on $D$ is expected to be particularly interesting. In Fig. \ref{fig1} we depict the dependence of the  rescaled diffusion coefficient $D/D_0$ on the ac-driving amplitude $a$, where $D_0$ corresponds to free thermal diffusion $D_0 = Q/\gamma$. One can notice there two distinctive features, namely, (i) this quantity exhibits giant damped quasiperiodic oscillations as $a$ increases and (ii) the maximal value of $D$ is several decades larger than the Einstein diffusion coefficient $D_0$.  E.g. for $\omega = 2$, the nonequilibrium diffusion coefficient is $D \approx 30 D_0$. Similar oscillations of diffusion has been observed in other systems. In Ref. \onlinecite{peter-epl}, the overdamped Brownian particle moves in  a sawtooth potential and is driven by a time-periodic piecewise constant driving force. The diffusion coefficient is also periodically damped with respect to the "tilting time" (the time when the driving force is non-zero) and the maximal amplification is $D \approx 14 D_0$. In Ref. \onlinecite{march-prl}, the overdamped limit of the dynamics (\ref{dimless-model}) was considered and the authors found oscillations of diffusion with its maximal value reading $D \approx D_0$. These results should be contrasted with giant variability of the latter quantity reported here in which it changes by two orders of magnitude for the full inertial dynamics characterized by the term $M\ddot x$ in Eq. (1).

Let us now explain the mechanism of the observed oscillatory diffusive behavior. For this purpose we first consider a fixed set of parameters with $\omega = 2$ for which this effect is the most prominent, see Fig. \ref{fig2}. We present there the rescaled diffusion coefficient $D/D_0$ (panel (a)) together with the variance $\sigma_{\overline{v}}^2$ (panel (b)) of the time averaged velocity as a function of the ac-driving amplitude $a$. Unexpectedly, the behavior of $D/D_0$ and $\sigma^2_{\overline{v}}$ is very similar. The subsequent maxima and minima are located at the same value of the amplitude, $a_1=9.8, a_2=15.3, a_3=22, a_4=28.2$. The reader can conclude that the magnitude of $D/D_0$ is strictly related to the spread of time averaged velocity measured by the variance $\sigma_{\overline{v}}^2$. Larger fluctuations of time averaged velocity lead to greater span of particle trajectories and consequently to superior diffusion coefficient. We note that the difference in $\sigma_{\overline{v}}^2$ corresponding to the first maximum ($a = 9.8$) and minimum ($a = 15.3$) is notable and equals almost two orders of magnitude.

In the next step we consider deterministic $Q = 0$ and noisy dynamics $Q = 0.5$ for two values of the ac-driving amplitude $a$ characterizing the first maximum ($a = 9.8$) and subsequent minimum ($a = 15.3$). In Fig. \ref{fig3} we present the corresponding probability distributions $P(\overline{v})$ for the time averaged velocity $\overline{v}$ of the Brownian particle.
In the deterministic case $Q=0$, there are two classes of trajectories in this parameter regime, namely, the running solutions (the motion is unbounded in space) and the locked solutions (the particle is trapped in the potential well). We note that for the noisy dynamics with $Q = 0.5$ fluctuations of time averaged velocity are much larger for the first maximum ($a = 9.8$)  than for the subsequent  minimum ($a = 15.3$). This observation confirms the result presented in Fig. \ref{fig2}. In left panels of Fig. \ref{fig3} we depict the same characteristics but now for the deterministic system with $Q = 0$. The reader can detect there the multistable dynamics: the locked states with $\overline{v} = 0$ coexist with a family of running ones $\overline{v} \neq 0$. Since the system given by Eq. (\ref{dimless-model}) is symmetric and the directed transport must vanish $\langle \overline{v} \rangle \equiv 0$ the running solutions can occur only in pairs $\overline{v} = \pm \overline{v}_i$.

%Non-zero thermal noise activates all possible random transitions between the coexisting states: locked $\to$ locked, running $\to$ running, locked $\to$ running and reversed.
A typical single trajectory of the noisy system with $Q>0$ contains: (i) extended periods when the particle is locked in one or several potential wells and (ii) time intervals when it is running. Therefore non-zero thermal noise activates transitions between the states coexisting for the deterministic system. For such multistable velocity dynamics there are three contributions to spread of trajectories of the system and consequently to the diffusion coefficient $D$. The first, which is the leading one,  comes from the spread associated with relative distance between the locked and running trajectories. The second and third parts correspond to thermally driven spread of trajectories following the locked and running solutions, respectively. The magnitude of these three contributions are related to stationary probabilities $p_0$ and $p_r$ of the particle to reside onto the locked and running trajectories, respectively. In the noisy case with dimensionless temperature $Q=0.5$ (which is rather "high" temperature) the given particle trajectory is classified as the locked if the corresponding time average velocity $\overline{v} \approx 0$. In the case $\overline{v} \neq 0$ it is categorized as the running one. It allows to reformulate the diffusion coefficient $D$ in terms of them. It is expected to be maximal when the share of the first contribution associated with the spread between the locked and running solutions is peaked.
\begin{figure}[t]
	\centering
	\includegraphics[width=0.9\linewidth]{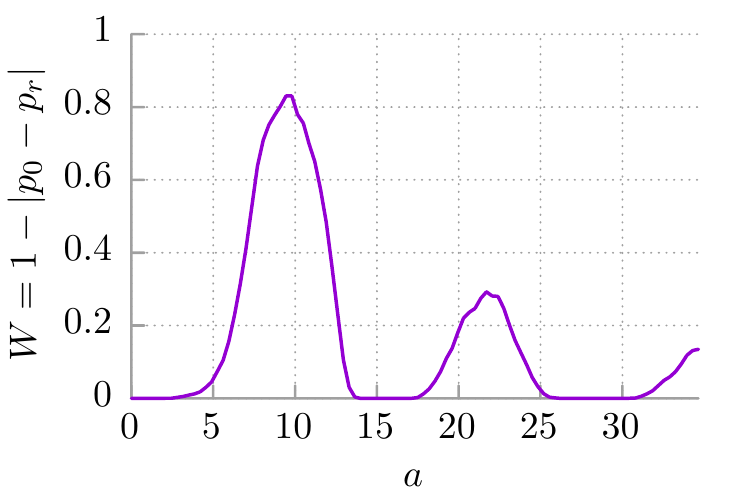}
	\caption{The quantifier $W = 1 - |p_0 - p_r|$ characterizing the difference in number of locked and running trajectories is depicted as a function of the ac-driving amplitude $a$. Other parameters are the same as in Fig. \ref{fig2}.}
	\label{fig4}
\end{figure}
To this aim, let us define the quantifier $W = 1 - |p_0 - p_r|$ which describes the difference in 
number of locked and running trajectories. If $W \neq 0$ then these two classes of solutions coexist. When $p_0 = p_r$ the quantifier is maximal $W = 1$. It corresponds to the situation in which locked and running trajectories are equiprobable. In Fig. \ref{fig4} we depict $W$ as a function of the ac-driving amplitude $a$.
%The given particle trajectory was classified as the locked solution if the corresponding time averaged velocity $\overline{v}$ obeyed the inequality $|\overline{v}| < 0.015$. In the opposite case it was categorized as the running solution.
The reader can observe that the maxima of this curve are exactly at the same points as the corresponding ones for the diffusion coefficient $D$, c.f. Fig. \ref{fig1} for $\omega = 2$. Moreover, the second maximum is significantly smaller than the first one and this fact is translated into the minor second maximum in the dependence of diffusion coefficient $D$. Therefore we can conclude that the giant damped quasiperiodic diffusion oscillations observed in Fig. \ref{fig1} can be characterized by the quantifier $W$ which describes the share of the spread of trajectories coming from the relative distance between the locked and running trajectories.

\section{Conclusions}
In this work we revisited the problem of diffusion in driven periodic system. As a particular instance we picked a paradigmatic model of the Brownian particle dwelling in a periodic potential and subjected to an external time periodic force. Since the latter takes the analyzed system out of equilibrium it is expected to possess the impact on its diffusive properties. Indeed, in this work we detected a peculiar effect of giant damped quasiperiodic diffusion oscillations controlled by the amplitude of external driving force. As a mechanism of this phenomenon we identified the corresponding oscillations of difference in the number of of locked and running solutions which carries the leading contribution to the spread of system trajectories, time averaged velocities and consequently also to the diffusion coefficient.

Our findings can be viewed as another manifestation of the impact of the dynamics in the velocity subspace onto the kinetics in the coordinate subspace. Recently it has been discovered that similar interplay may be also \emph{modus operandi} of fluctuation-induced dynamical localization \cite{spiechowicz2017scirep, spiechowicz2019chaos} causing anomalous diffusion in systems which at the first glance should not react in this way. The problem how the directed transport influences the diffusive behavior is largely unexplored and therefore we expect new instances of it to appear in the near future, including also the experimental ones. It is facilitated by the fact that the considered system has a multitude of physical realizations \cite{risken} with Josephson junctions and cold atoms dwelling in optical lattices.    E.g., for a system consisting of a resistively and  capacitively shunted Josephson junction device, 
the coordinate $x(t)$ corresponds to the phase
difference $\phi(t)$  between the macroscopic wave functions of the
Cooper pairs on both sides of the Josephson junction and the velocity $v(t)$ corresponds to  the voltage $V(t)$ across the junction. Because it would be difficult to measure diffusion of the phase instead of it experimentalists can measure the voltage fluctuations from which they can extract information on  diffusion of the phase.  Finally, our  findings can be used for control of diffusion according to the procotol presented in Fig. \ref{fig2}: if  large diffusion is desired one has to find a value of the amplitude $a$ (here $a=9.8)$ which maximizes $D$. In turn, if  one needs to reduce diffusion the suitable value of  $a$ (here $a=15.3$) should be matched to minimalize the diffusion coefficient. Moreover, at the maximum 
$D_{max} \approx 30 D_0 > D_0$ and at the minimum $D_{min} \approx 0.3 D_0 < D_0$, i.e. $D_{max} \approx 100 D_{min}$ and one can control diffusion over the range of two orders  of magnitude.  

\section*{Acknowledgement}
This work was supported by the Grants No. NCN 2017/26/D/ST2/00543 (J. S), No. NCN  2018/30/E/ST3/00428 (I. G. M.) and in part by PLGrid Infrastructure.  I. G. M. acknowledges University of Silesia for hospitality  since 24 February 2022.  

%\section*{Appendix}
%Here, for comparison with the previous case considered for the first pair of maximum ($A=10$) and minimum (A=15), we present histograms of locked and running states corresponding  to the second pair of maximum $A=22$ and minimum $A=27$ of the diffusion coefficient. 
%We depict both  the deterministic  and noisy  asymptotic velocity distribution. 
%Again, as in the previous case presented in Fig. \ref{fig3}, we observe  that for noisy dynamics (when intensity of thermal noise is $Q=0.5$), fluctuations of stationary velocity in  the case $A=22$ are much larger than for $A=27$.  Larger fluctuations (variance) of velocity lead to larger spread of trajectories and in consequence to the larger diffusion coefficient.    
%Moreover, in the considered regime and  in the absence of thermal noise, bistable dynamics is observed: there are locked and running solutions.
%
%\begin{figure}[t]
%	\centering
%	\includegraphics[width=0.48\linewidth]{max2.png}
%	\includegraphics[width=0.48\linewidth]{max22.png}
%	\includegraphics[width=0.48\linewidth]{min2.png}
%	\includegraphics[width=0.48\linewidth]{min22.png}
%	\caption{Left panel: Histograms of locked and running solutions at temperature $T=0$ (left panels)  and $T=0.5$ (right panels). 
%	Upper panels are for the second maximum  of $D$ at $A=22$ and bottom panels are for the second minimum at $A=27$.  }
%	\label{fig3}
%\end{figure}
%

\end{document}